\documentstyle[twoside,fleqn,espcrc2]{article}
\input{epsfig.sty}

\newcommand{\AmS}{{\protect\the\textfont2
  A\kern-.1667em\lower.5ex\hbox{M}\kern-.125emS}}
\def\beq{\begin{equation}}
\def\eeq{\end{equation}}
\def\beqa{\begin{eqnarray}}
\def\eeqa{\end{eqnarray}}
\def\MeV{\nobreak\,\mbox{MeV}}
\def\GeV{\nobreak\,\mbox{GeV}}

\def\bb{\scriptsize\mbox{b }}

\title{Hyperon-nucleon coupling from QCD sum rules}
\author{M.E. Bracco
\address{Instituto de F\'{\i}sica,
        Universidade do Estado do Rio de Janeiro, \\
        Rua S\~ao Francisco Xavier 524 - 20559-900, Rio de Janeiro, RJ, Brazil}%
        , F.S. Navarra $^{\bb}$ and
        M. Nielsen
\address{Instituto de F\'{\i}sica, 
        Universidade de S\~{a}o Paulo, \\
        C.P. 66318,  05389-970 S\~{a}o Paulo, SP, Brazil}}

\begin{document}

\begin{abstract}
The $NKY$ coupling constant for $ Y = \Lambda $ and $ \Sigma $ 
 is evaluated in a QCD sum rule calculation. 
We discuss and extend the result of a previous 
analysis in the $\rlap{/}{q}i\gamma_5$  structure and compare it with the 
result obtained with the use of the $\gamma_5 \sigma_{\mu \nu}$ structure. 
We find a huge violation of the SU(3) symmetry
in the $\gamma_5 \sigma_{\mu \nu}$ structure.

\end{abstract}

\maketitle

In understanding the dynamics of kaon-nucleon scattering or the strangeness
content of the nucleon using hadronic models, it is important to know the
hadronic coupling constants involving the kaons. Among them, $g_{NK\Lambda}$ and
$g_{NK\Sigma}$ are the most relevant coupling constants. To determine these
couplings using the QCD sum rules \cite{svz} one can follow two different
approaches: a) the two-point function, where the nucleon and hyperon fields
are sandwiched between the vaccum and kaon states, or b) the three-point
function where three interpolating fields are sandwiched between vacuum states.

In the case of the pion-nucleon coupling constant, in a pioneer calculation 
\cite{rry} both approaches showed to reproduce the phenomenological value fairly
well. However, in this first study the continuum contribution was neglected and 
since then many calculations were done including higher order terms in
the operator product expansion (OPE) and the continuum contribution \cite{sh}, 
going beyond the soft-pion
limit and including also pole-continuum transitions \cite{bk}.

For the nucleon-kaon-hyperon coupling constant there are also QCD sum rules
calculations based on the two- and three-point functions \cite{cch,bnn,as}. The
advantages of the three-point function calculation is that it allows for the
calculation of the form factors at the hadronic vertices. 

In the strange sector, the nucleon-kaon-hyperon form factors are used, for 
instance, to evaluate  the strange radius of the nucleon using the kaon cloud
\cite{fnjc} and, therefore, a theoretically founded evaluation of these form
factors is wellcome.

We will calculate the $g_{NKY}$ coupling constant using the three-point function
\beqa
A(p,p^\prime,q)&=&\int d^4x \, d^4y \,  
\, e^{ip^\prime x} \, e^{-iqy}
\nonumber\\
&\times&\langle 0|T\{\eta _{Y}(x)
j_5(y)\overline{\eta }_N(0)\}|0\rangle 
\label{cor}
\eeqa
where $j_5=\bar{s}i\gamma_5u$.

	As it is well known from two-point sum rules for baryons, there is a 
continuum of choices for the baryon interpolating fields. Of course the results
should be independent of the choice of the current, if we considered an infinity 
number of terms in the OPE and if we had a perfect model for the continuum
contribution in the phenomenological side. However, the OPE has to be truncated 
and we work  with a very simple model for the continuum contribution. Therefore,
the results do depend on the choice of the currents. For the proton $\Lambda$
and $\Sigma$ we can write general currents as  \cite{dosch}
\beq
{\eta }_{P}= 2
\varepsilon _{abc}[({u}_a^TC{d}_b)\gamma_5 u_c 
+b ({u}_a^TC\gamma _5{d}_b)u_c]\;,
\label{lambcur}
\eeq
\beq
\eta_{Y}=2[\eta_{Y_1}+b\eta_{Y_2}]\;,
\label{lambhy}
\eeq
where $b$ is a parameter and
\beqa
{\eta }_{\Lambda_1}&=& {1\over\sqrt{6}}
\varepsilon _{abc}[2({u}_a^TC{d}_b)\gamma_5 s_c +({u}_a^TC{s}_b)\gamma_5 d_c 
\nonumber\\
&-& ({d}_a^TC{s}_b)\gamma_5u_c]\;,
\eeqa
\beqa
{\eta }_{\Lambda_2}&=& {1\over\sqrt{6}}
\varepsilon _{abc}[2({u}_a^TC\gamma_5{d}_b) s_c+ ({u}_a^TC\gamma_5{s}_b) d_c 
\nonumber\\
&-& ({d}_a^TC\gamma_5{s}_b)u_c]\;,
\eeqa
\beq
{\eta }_{\Sigma_1}= {1\over\sqrt{2}}
\varepsilon _{abc}[ ({u}_a^TC{s}_b)\gamma_5 d_c +
 ({d}_a^TC{s}_b)\gamma_5u_c]\;,
\eeq
\beq
{\eta }_{\Sigma_2}= {1\over\sqrt{2}}
\varepsilon _{abc}[ ({u}_a^TC\gamma_5{s}_b) d_c +
 ({d}_a^TC\gamma_5{s}_b)u_c]\;.
\eeq
In ref.~\cite{dosch} it was shown that the best interpolating fields
for mass sum rules have $b=-1/5$. However, to be able to compare our results
with previous calculations we will use $b=-1$.

In the phenomenological side the  matrix element of the pseudoscalar current 
between the hyperon and nucleon states defines the pseudoscalar 
form-factor
\begin{equation}
\langle Y(p^\prime)|j_5|N(p)\rangle=g_P(q^2)\overline{u}(p^\prime)i
\gamma_5
u(p)\; ,
\label{ps}
\end{equation}
where $u(p)$ is a Dirac spinor and $g_P(q^2)$ is related to $g_{NDY}$
through the relation \cite{rry}
\beq
g_P(q^2)={m_K^2 f_K\over m_q}{g_{NKY}\over q^2-m_K^2} \; ,
\label{g}
\eeq
where $m_K$ and $f_K$ are the kaon mass and decay constant and
$m_q$ is the average of the quark masses: $(m_u+m_s)/2$.

Saturating the correlation function Eq.(\ref{cor}) with
$Y$ and $N$ intermediate states, and using Eq.~(\ref{g})we get 
\beqa
&&A^{(phen)}(p,p^\prime,q) =\lambda_{Y}\lambda_N {m_K^2 f_K
\over m_q}{g_{NKY}\over q^2-m_K^2}
\nonumber\\
&&
\times{1\over p'^2-M_{Y}^2}{1\over p^2-M_N^2}\left[(M_{Y}M_N-p.
p^\prime)i\gamma_5\right.
\nonumber\\
&&
+ {M_{Y}+M_N\over2}\rlap{/}{q}i\gamma_5 - (M_{Y}-M_N)
\rlap{/}{P}i\gamma_5
\nonumber\\
&&
-\left.\sigma^{\mu\nu}\gamma_5q_\mu p^\prime_\nu \right] + 
\mbox{higher resonances}\; ,
\label{ficor}
\eeqa
where $\lambda_{Y}$ and $\lambda_N$ are
the couplings of the currents with the respective hadronic states.

In. Eq.~(\ref{ficor}) we clearly see four distinct Dirac structures.
In principle, any of these structures can be used to calculate $g_{NKY}$
and the sum rules should yield similar results. However, each sum rule
may have uncertainties due to the truncation of the OPE side and different 
contributions from the continuum. Therefore, depending on the Dirac structure
we can obtain different results due to these uncertainties. The traditional way
to control these uncertainties, and to check the reability of the sum rule,
is to choose an appropriate Borel window.

In a recent evaluation of the pion-nucleon coupling constant \cite{klo} it was
pointed out that the structure $\gamma_5\sigma_{\mu\nu} q_\mu p^\prime_\nu $ 
gives a better result
since it gets a smaller contribution from the single pole term coming
from $N\rightarrow N^*$ transition and it is also independent of the effective 
model employed in the phenomenological side. Therefore, we will write a sum 
rule for the  $\sigma^{\mu\nu} \gamma_5 q_\mu 
p^\prime_\nu$ structure. As we are interested in the value of the
coupling constant at $q^2=0$, we will make a Borel transform to
both $p^2={p^\prime}^2\rightarrow M^2$. In Eq.~(\ref{ficor}) higher
resonances refers to pole-continuum transitions as well as pure continuum 
contribution. The pure continuum contribution 
will be taken into account as usual through the standard form
of ref.\cite{ioffe}.
 
In the OPE side only even dimension operators contribute to the 
$\sigma^{\mu\nu} \gamma_5 q_\mu p^\prime_\nu$ structure, since the dimension 
of Eq.(\ref{cor})
is four and $q_\mu p^\prime_\nu$ take away two dimensions.
We will neglect $m_K^2$ and $ m_s^2$ in the 
denominators and, 
consequently, only terms proportional to $ 1/ q^2$ will contribute to the sum 
rule. We will consider diagrams up to dimension six.

To evaluate the continuum contribution we write a double dispersion relation
for the invariant function corresponding to the $\sigma^{\mu\nu} 
\gamma_5 q_\mu p^\prime_\nu$ structure:
\beq
F(P^2,{P^\prime}^2,Q^2)=\int ds ds^\prime{\rho(s,s^\prime,Q^2)\over(s+P^2)
(s^\prime+{P^\prime}^2)}\;.
\eeq

The function $\rho(s,s^\prime,Q^2)$ (which is proportional to the
double discontinuity of $F$) can generally be written as
\beqa
&&\rho(s,s^\prime,Q^2)=a(Q^2)\delta(s-M_N^2)\delta(s^\prime-M_Y^2)+
\nonumber\\
&&b(s^\prime,Q^2)\delta(s-M_N^2)+c(s,Q^2)\delta(s^\prime-M_Y^2)+
\nonumber\\
&&\rho^{\scriptsize\mbox{OPE}}(s,s^\prime,Q^2)\theta(s-s_N)
\theta(s^\prime-s_Y)\;,
\label{esp}
\eeqa
where the last term is the tradicional way of taking into account
the pure continuum contribution with $s_N$ and $s_Y$ being the
continuum thresholds for the nucleon and hyperon respectively. The first term
in Eq.~(\ref{esp})
gives the double pole contribution and the second and third terms correspond
to single pole contributions coming from pole-continuum transitions. 
In ref.~\cite{is} it was shown that single pole contributions are not 
suppressed by the Borel transformation and give rise to terms proportional
to $M^2$. Therefore, to take into account the single pole term in the Borel 
tranformed sum rule we make the substitution:
\beq
g_{NKY}\rightarrow g_{NKY} +AM^2\; ,
\eeq
where $A$ is an unknown parameter. The sum rules we get for $g_{NK\Lambda}$ and
$g_{NK\Sigma}$ are \cite{bnn}
\beqa
&&g_{NK\Lambda}+AM^2=
\nonumber \\
&&-\sqrt{2\over3}{1\over\tilde{\lambda}_{\Lambda}\tilde{\lambda}_N}{m_q\over 
m_K^2 f_K}{M_{\Lambda}^2-M_N^2\over e^{-M_N^2/M^2}-e^{-M_{\Lambda}^2/M^2}}\
\nonumber \\
&&\times
a\left[{m_s\gamma\over4}M^2E_0^{\Lambda}
-{4\over3}a(1+\gamma)\right] \; ,\label{sr}
\eeqa
\beqa
&&g_{NK\Sigma}+BM^2=
\nonumber \\
&&\sqrt{2}{1\over\tilde{\lambda}_{\Sigma}\tilde{\lambda}_N}{m_q\over 
m_K^2 f_K}{M_{\Sigma}^2-M_N^2\over e^{-M_N^2/M^2}-e^{-M_{\Sigma}^2/M^2}}\
\nonumber \\
&&\times{am_s\gamma\over4}M^2E_0^{\Sigma} \; .\label{srsi}
\eeqa
where $a=-(2\pi)^2 \langle\overline{q}q\rangle \simeq 0.5 \, \GeV^3$,  
$\gamma = \langle\overline{s}s\rangle/ \langle\overline{q}q\rangle  
\simeq 0.8$ and $\tilde{\lambda}_B=(2\pi)^2\lambda_B$. In the above expressions 
$E_0^{Y}=1-e^{-s_{Y}/M^2}$, and this factor
in Eqs.~(\ref{sr}) and (\ref{srsi}) accounts for the continuum contribution.
The parameters $A$ and $B$
denotes the contribution from the unknown single pole term coming from
$N\rightarrow N^*$ transition. 

In this calculation 
the continuum  thresholds are chosen to be  
$s_{Y}\,=\,(M_{Y}+0.5)^2\,\GeV^2$. The hadron 
masses are $M_N\,=\,0.938\,\GeV$, $M_{\Lambda}\,=\,1.150\,\GeV$, 
$M_{\Sigma}\,=\,1.189\,\GeV$ and 
$m_K\,=\,0.495\,\GeV$. The strange quark mass is taken to be $m_s\,=\,150$ 
\, MeV and the kaon decay constant is $f_K={160\over\sqrt{2}}\MeV
\simeq113\MeV$.                    
The relevant Borel mass here is $M\simeq\frac{M_N+M_{\Lambda}}{2}$ and 
we analyse the sum rule in the interval $0.8 \leq M^2 \leq 1.6$ 
\GeV$^2$ where the continuum contribution is always smaller than 50\% of
the total OPE.
\begin{figure}[htb]
\centerline{\epsfig{figure=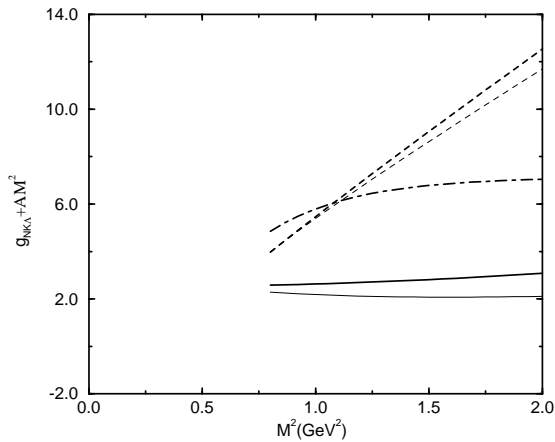,width=8cm}}
\vspace{-1cm}
\caption{\small{$g_{NK\Lambda}$ as a function of the squared Borel mass
$M^2$ for the $\gamma_5\sigma_{\mu\nu}$ structure (solid line) and for the
$\rlap{/}{q}\gamma_5$ structure with (dashed line) and without (dot-dashed 
line) continuum contibutions. The thick lines are obtained using the continuum
thresholds given by: $s_{\Lambda}\,=\,(M_{\Lambda}+0.5)^2\,\GeV^2$,
while for the thin lines lines we used
$s_{\Lambda}\,=\,(M_{\Lambda}+0.7)^2\,\GeV^2$.}}
\label{fig1}
\end{figure}

In Figs.~1 and 2 we show the RHS of Eqs.~(\ref{sr}) and (\ref{srsi})
as a function of the Borel
mass squared (thick solid line). We show the results in a broader Borel
range than discussed above to show that our conclusions are not very 
constrained by the Borel window used. To check the sensitivity of our result 
on the continuum contribution, we have increased the continuun thresholds
as: $s_{Y}\,=\,(M_{Y}+0.7)^2\,\GeV^2$ and 
 plotted the corresponding result as the 
thin line in the same figures. As a first sign it seems that the result is
very sensitive to the continuum thresholds. However, as the value of the
coupling constant is obtained by the extrapolation of the line to $M^2=0$,
we imediately see that both curves lead to approximately the same result. 
Indeed, fitting the QCDSR result to a straight line we get

$$
{|g_{NK\Lambda}|} \,=\,2.4\,\pm\,0.1 \;\,;\;
{|g_{NK\Sigma}|} \,=\,0.03\,\pm\,0.02\; ,
$$
where the errors were estimated by using the two different thresholds.
The values of $A$ and $B$ are very small showing that
the single pole contribution is not very important in this structure, in
agreement with the results in ref.\cite{klo}.
\begin{figure}[htb]
\centerline{\epsfig{figure=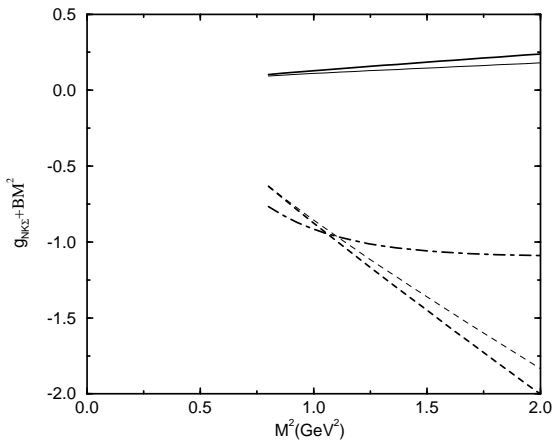,width=8cm}}
\vspace{-1cm}
\caption{\small{Same as Fig.~1 for $g_{NK\Sigma}$.}}
\label{fig2}
\end{figure}
The results obtained in \cite{cch} are :
${|g_{NK\Lambda}|} \,=\,6.96$ and ${|g_{NK\Sigma}|} \,=\,1.05$.
However, the results in ref.\cite{cch} were obtained without considering 
continuum contribution and they are shown as the dot-dashed lines in Figs.~1
and 2. Once the continuum contribution is included, through the usual $E_i$ 
factors, the behaviour of the sum rule as a function of the Borel mass 
changes drastically, as can be seen by the dashed line in Figs.~1 and 2.
In particular, both  
$g_{NK\Lambda}$ and  $g_{NK\Sigma}$ become approximately linear functions 
of $M^2$, showing the importance of the pole-continuum  
contribution in this structure. 

Fitting the RHS  of the sum rule results on the structure $\rlap{/}{q}i
\gamma_5$ \cite{cch} (including the 
continuum contribution) to a straight line one gets
$$
{|g_{NK\Lambda}|}=1.5\pm0.3\,\,;\,
{|g_{NK\Sigma}|}=0.25\pm0.05\; ,
$$
where the errors are again evaluated by considering the two different
continuum thresholds.

As in ref.\cite{klo} we find out that we can obtain very different results 
for the coupling constants depending on the structure considered. Of
course the procedure used here to extract the coupling constant (fitting the
QCDSR result to a straight line in a given Borel window and extrapolating it
to $M^2=0$) is more reliable when the single pole term is small. Therefore,
the results obtained for the structure $\rlap{/}{q}i\gamma_5$ may contain big 
errors since the single pole contribution to this structure is very strong, as
can be seen by the dashed lines in Figs.~2 and 3. On the other hand, we may
say that the results on the structure $\gamma_5 \sigma_{\mu \nu}$, 
analysed here, are not contaminated by the single pole transitions and its
extraction with the method used here is more reliable.

As a final remark we note that the values for the coupling constants obtained 
here in both structures considered, are not in agreement with the
exact SU(3) symmetry. In this limit one gets \cite{bnn} 
$|g_{NK\Lambda}/g_{NK\Sigma}|=3.55$. Therefore, our
results show a huge breaking of SU(3) symmetry. One important question is 
if this breaking of the SU(3) symmetry is related with the particular choice
of the interpolating fields. Work in this direction is in progress \cite{dnnn}.

\end{document}